\documentclass[sigconf]{acmart}

\usepackage[utf8]{inputenc} 
\usepackage[T1]{fontenc}    
\usepackage{hyperref}       
\usepackage{url}            
\usepackage{booktabs}       
\usepackage{amsfonts}       
\usepackage{nicefrac}       
\usepackage{microtype}      
\usepackage{lipsum}
\usepackage{graphicx}
\usepackage[most]{tcolorbox}
\usepackage{caption}
\usepackage{subfigure}
\usepackage{relsize}
\usepackage{enumitem}
\usepackage{titlesec}
\titlespacing{\section}{0pt}{4pt plus 1pt minus 1pt}{0pt plus 1pt minus 1pt}
\titlespacing{\subsection}{0pt}{4pt plus 1pt minus 1pt}{0pt plus 1pt minus 1pt}
\titlespacing{\subsubsection}{0pt}{2pt plus 1pt minus 1pt}{0pt plus 1pt minus 1pt}

\hypersetup{
    colorlinks=true,
    linkcolor=red,
    anchorcolor=black,
    filecolor=magenta,      
    urlcolor=blue,
    pdftitle={Sharelatex Example},
    bookmarks=true,
    pdfpagemode=FullScreen,
}

\AtBeginDocument{%
  \providecommand\BibTeX{{%
    \normalfont B\kern-0.5em{\scshape i\kern-0.25em b}\kern-0.8em\TeX}}}

\setcopyright{none}
\renewcommand\footnotetextcopyrightpermission[1]{}

\begin{document}
\pagestyle{plain}
\title{Don’t cross that stop line: Characterizing Traffic Violations in Metropolitan Cities}



\author{Shashank Srikanth}
\authornote{These authors contributed equally to work.}
\affiliation{%
  \institution{IIIT Hyderabad}
}
\email{shashank.s@research.iiit.ac.in}

\author{Aanshul Sadaria}
\authornotemark[1]
\affiliation{
\institution{IIIT Hyderabad}}
\email{aanshul.sadaria@students.iiit.ac.in}

\author{Himanshu Bhatia}
\authornotemark[1]
\affiliation{
\institution{IIIT Hyderabad}}
 \email{himanshu.bhatia@students.iiit.ac.in}

\author{Kanay Gupta}
\authornotemark[1]
\affiliation{
    \institution{IIIT Hyderabad}
}
\email{kanay.gupta@students.iiit.ac.in}

\author{Pratik Jain}
\authornotemark[1]
\affiliation{%
  \institution{IIIT Hyderabad}}
\email{pratik.jain@students.iiit.ac.in}

\author{Ponnurangam Kumaraguru}
\authornote{Major part of the work was done during a year long sabbatical at IIIT Hyderabad.}
\affiliation{ 
  \institution{IIIT Delhi}}
\email{pk@iiitd.ac.in}

\begin{abstract}
\vspace{1.0em}
In modern metropolitan cities, the task of ensuring safe roads is of paramount importance. Automated systems of e-challans (Electronic traffic-violation receipt) are now being deployed across cities to record traffic violations and to issue fines. In the present study, an automated e-challan system established in Ahmedabad (Gujarat, India) has been analyzed for characterizing user behaviour \footnote{Note that we have used the term "user". We assume that each vehicle is associated with a unique individual and henceforth will consider them to be equivalent.}, violation types as well as finding spatial and temporal patterns in the data. We describe a method of collecting e-challan data from the e-challan portal of Ahmedabad traffic police and create a dataset of over $3$ million e-challans. The dataset was first analyzed to characterize user behaviour with respect to repeat offenses and fine payment. We demonstrate that a lot of users repeat their offenses (traffic violation) frequently and are less likely to pay fines of higher value. Next, we analyze the data from a spatial and temporal perspective and identify certain spatio-temporal patterns present in our dataset. We find that there is a drastic increase/decrease in the number of e-challans issued during the festival days and identify a few hotspots in the city that have high intensity of traffic violations. In the end, we propose a set of $5$ features to model recidivism in traffic violations and train multiple classifiers on our dataset to evaluate the effectiveness of our proposed features. The proposed approach achieves $95\%$ accuracy on the dataset. 
\end{abstract}






\maketitle
\thispagestyle{empty}

\section{Introduction}
Traffic accidents were responsible for over $1$ million deaths all over the world in the year 2016 \cite{deathcauses}. Of these accidents, more than $90\%$ occur in developing countries. Previous research in the field of behaviour studies regarding traffic rule violations has shown that in greater than $70\%$ of the cases, the role of human behaviour is one of the causes \cite{jha2017traffic}. Most of these accidents can be prevented if the traffic rules are properly followed and as a result, the traffic police across states in India are adopting automated traffic management systems to promote adherence to traffic rules \cite{automatedsystem}. These automated systems are capable of varied tasks like capturing violations and issuance of e-challan (Electronic traffic-violation receipts) without any human intervention. These systems can also generate e-challans along with photo evidence and send it to violators through SMS/email/post. 
\begin{figure}[b!]
    \centering
    \includegraphics[scale=0.35]{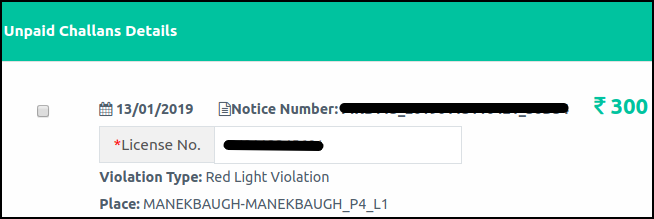}
    \caption{A sample e-challan. It contains several important details regarding the traffic violations such as the time and location of violation along with the type of violation and corresponding fine amount. It also includes data about whether the e-challan has been paid or not. The license number and notice number have been hidden to protect privacy.}
    \label{fig:sample-challan}
\end{figure}
Figure \ref{fig:sample-challan} shows an example of an e-challan generated in Ahmedabad. The e-challan consists of several details of a given traffic violation like the time, place of violation and other information like the violation type and corresponding fine amount. Thus, automated traffic management systems like those in Ahmedabad can be mined to extract traffic violations data and used for estimating the possibility of repeat offenses. Such datasets can also be used to characterize the type of traffic violations in a city and identify the spatial and temporal patterns of the traffic violations. The Ahmedabad traffic police launched their automated traffic management system in 2015 and it leverages a network of $6$,$000$ video surveillance cameras (dedicated to detect \textit{red light violations}) installed across $130$ traffic junctions. The system has been an enormous success and a total of $1.27$ million \textit{stop line violation} e-challans were issued in the year 2018 \cite{eChallan}. 


Despite the ever-increasing amount of traffic violations data being available, there has not been a systematic analysis of such violations. Such an analysis would be particularly useful for the government which is responsible for framing laws and the police which is responsible for making the roads safer for citizens. In this work, we carry out a longitudinal study of e-challan receipts in the city of Ahmedabad and investigate the effectiveness of the above system in reducing repeat offences. Unlike some earlier works \cite{choudhary2015identification, ren2018deep}, which analyze road accident data, we restrict our analysis only to traffic violations. In order to understand traffic violations from the prism of big-data analysis, we address the following questions.

\subsection{Research Questions}
Different types of traffic violations require different varieties of preventive measures. Similarly, not all people are equally prone to committing traffic violations, and quite a few of them can be serial offenders. Unlike traditional challan systems, where payments were made at the time of issue itself, e-challan systems allow people to pay the fines later. Thus, people often delay paying their fines unless forced to or are simply unaware of the fines being issued to them \cite{ahmedabadchallan}. Therefore, we ask our first research question:

\textbf{RQ1: } \textit{What are the characteristics of the users who are issued the e-challans and their corresponding traffic violations? }

The e-challan issue date and fine payments date exhibit some interesting temporal patterns. Besides, the information from the spatial dimension will also be useful to clusters of regions in the city where violations are more likely to occur. Thus, we are interested in solving:

\textbf{RQ2: } \textit{What are the temporal and spatial patterns of traffic violations?}

A lot of research has been conducted on recidivism cases amongst prisoners and sex offenders \cite{hanson1998predicting, harris1991psychopathy}. Most of these works focused on identifying factors that are strongly related to recidivism. Using similar methods to predict recidivism in the case of traffic violations would be very useful for traffic police and government authorities. Thus, we propose the following research question:

\textbf{RQ3: } \textit{Can instances of repeat offences be predicted with reasonable accuracy based on just the history of traffic violations of a person?}

\subsection{Contributions}
The main contributions of this work are:
\begin{itemize}[noitemsep]
    \item We describe a method to collect large-scale data of traffic violations from the e-challan portal of Ahmedabad and collect a dataset of over $3$ million e-challans.  
    \item We characterize the user behaviour in our dataset with respect to repeat offences and fine payment. We show that users with more violations are less likely to pay their fines and users, in general, prefer to pay the smaller fines as compared to the larger ones. Moreover, we also show that a few of the violations account for most of the e-challans. 
    \item We analyze the temporal distribution of the traffic violation data over all the $4$ years and show that there is a drastic increase/decrease in the number of fines issued during the festival days. And we show the emergence of new traffic violation hot-spots over time by performing spatial and temporal analysis simultaneously.
    \item We create a dataset consisting of over $600$,$000$ users and their corresponding recidivism history. We identify and describe several attributes that can be inferred from a user's past violation history and be used for predicting recidivism. 
\end{itemize}

\subsection{Privacy and Ethics} 
We collect data from Ahmedabad traffic police's e-challan portal \footnote{\hyperlink{https://payahmedabadechallan.org}{https://payahmedabadechallan.org}} and all the data used is publicly available. Additionally, we do not use any personally identifiable information for our analysis.
\linebreak
\linebreak
The rest of this paper is organized as follows: We discuss the related work in Section \ref{sec:relatedwork} and Section \ref{sec:dataset} discusses in detail the aspects of data collection. We characterize the user behaviour and the type of violations in our dataset in Section \ref{sec:characterizingusers} and in Section \ref{sec:spatiotemporal}, we identify the spatial and temporal patterns of traffic violations. We describe the proposed features and machine learning methods to predict recidivism in Section \ref{sec:recidivism}. The work is concluded in Section \ref{sec:discussion} and Section \ref{sec:futurework} deals with possible extensions of the project.

\section{Related Work}
\label{sec:relatedwork}
We structure the discussion of related work into two main themes: related work on traffic and road violations data, and work concerning predicting recidivism amongst convicts.

\textbf{Traffic Accidents and Violations}: Traffic accidents account for more than a million deaths each year across the world according to WHO \cite{trafficaccidents} and a lot of research has been conducted on data concerning road accidents. Sanjay et al. \cite{singh2017road} analyze the road accidents data across India at a national, state and metropolitan city level and show that distribution of road accident deaths and injuries vary according to age, gender, month and time. They also show that more than $50\%$ of the cities face higher fatality risks as compared to their rural counterparts. 
Another set of approaches to model road accidents involve identifying road accident hotspots using GIS (Geographical Information System) technologies. Choudhary et al.\cite{choudhary2015identification} geocoded $5$ years of road accident locations over the digital map of Varanasi and clustered accidents using a spatial heatmap. In contrast to these approaches, which analyze and model road accidents data, we analyze only the traffic violations data of the city of Ahmedabad. The insights gained from analyzing traffic violations would allow the appropriate agencies to take suitable preventive measures. Qiqi et al. \cite{wang2015common} show that traffic violations amongst bus drivers are associated with the date, weather, and presence of traffic cameras at bus stations. The analysis done by Qiqi et al. \cite{wang2015common} is the most similar to our work, but their dataset size is much smaller than ours in terms of size, and they do not focus on predicting recidivism.

\textbf{Predicting Recidivism}: A lot of research has been done on the task of predicting recidivism amongst convicts of crime. There have been several approaches to predicting recidivism which range from using survival time models \cite{schmidt1989predicting} and machine learning methods \cite{ritter2013predicting}. Work by Gerald et al. \cite{stahler2013predicting} proposes several independent characteristics like demographics, offense type, location and spatial contagion to model recidivism. They use GIS (Geographic Information System) and logistic regression modelling to show that the likelihood of re-incarceration was increased with male gender, offense type and certain locations. Random forest models have also been applied successfully to predict recidivism in \cite{ritter2013predicting} by leveraging a large number of features. Our work differs from all these methods as it aims to predict the recidivism of traffic violations, unlike other criminal activities. Similar to the work by Nancy et al. \cite{ritter2013predicting}, we also utilize a random forest model to predict recidivism of users based on their past violation history. 

\begin{figure}[h!]
    \centering
    \includegraphics[scale=0.27]{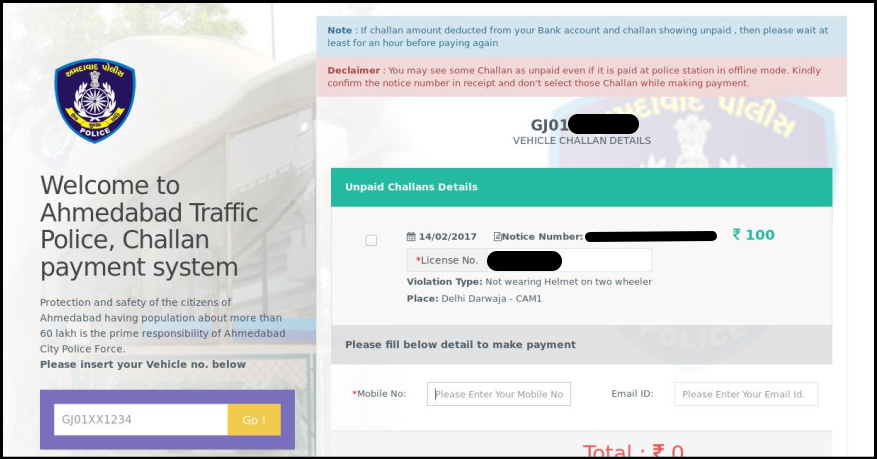}
    \vspace{-1.0em}
    \caption{Ahmedabad traffic police's e-challan portal - on entering the vehicle registration number in the bottom left text box, the details of unpaid e-challans are displayed in the right window pane.}
    \label{fig:echallan-portal}
\end{figure}

\section{Dataset collection and description}
\label{sec:dataset}

For this work, we collect traffic violations data (e-challan) from the e-challan portal of Ahmedabad traffic police \footnote{\hyperlink{https://payahmedabadechallan.org}{https://payahmedabadechallan.org}}. 

\subsection{Data collection}
Ahmedabad traffic police have an online \hyperlink{https://payahmedabadechallan.org}{e-challan portal} that is used by people to check the status of fines and also make the relevant payments online. A brief description of the portal usage and its features has been shown and discussed in Figure \ref{fig:echallan-portal}. The portal does not require a user to have a login ID or password, and one needs to enter their vehicle registration ID to get the details of all e-challans issued to that vehicle. This is in contrast to various other cities like Hyderabad, Mumbai, etc., where login ID with a password is mandatory. Thus, here we can view the set of e-challans issued to other vehicles as well. The portal does not reveal any personally identifiable information unless a user demands a receipt of payment. 

For obtaining the data through the e-challan portal, we leverage the Selenium headless web browser to make millions of web requests to their server and collect over $3$ million e-challans. In each new request, we provide a new vehicle registration number in the corresponding header files and obtain traffic violations data for that vehicle. As we do not have access to the list of registered vehicles in the city of Ahmedabad, we make a request to the portal with all possible vehicle registration numbers in Ahmedabad. Vehicle registration IDs across cities in India have a total of 10 characters and barring a few exceptions look like the following: \texttt{GJ-01-AB-1234}. The registration IDs are divided into four main parts:
\begin{itemize}
    \item First two letters indicate the state/union territory in which the vehicle is registered
    \item The next two numbers are sequential numbers of the registration district 
    \item The third portion represents the ongoing series of an RTO (Regional Transport Office) which issues the registration IDs
    \item The last part consists of a $4$ digit number, which combined with the previous three parts, is unique to each registered vehicle.
\end{itemize}

Thus, each vehicle is uniquely identified by its registration ID, and two vehicles cannot have the same ID. To get the violation data of all the vehicles in Ahmedabad, we enumerate all possible combinations of vehicle registration numbers that begin with the prefix of GJ01, which is one of the prefix codes for the city of Ahmedabad. We add a two-letter suffix and a $4$ digit number after the letters \texttt{GJ01} to get a total of $6$,$760$,$000 (26 * 26 * 10 ^ 4)$ vehicle registration IDs combinations and query the server with these IDs. 

One can also get the payment date of a paid e-challan by requesting for a receipt of the payment in the portal. Downloading and parsing millions of e-challan receipts in PDF format is computationally expensive, so we randomly sample $4$,$500$ users from our dataset that paid at least one of their fines and collect their fine payment data.  The payment receipt is provided in the form of a PDF, and we use \href{https://tabula.technology/}{\texttt{Tabula}} to parse the PDF into JSON format suitable for analysis.

\subsection{Dataset}
\label{sec:minidataset}

We collect a total of over $3$ million e-challans over a period of $4$ years from $2016-2019$. There were a total of $1$,$177$,$695$ unique vehicles with one or more e-challans and they together account for $808$,$004$,$725$ in fines to the government. For each unique e-challan in our dataset, we have five major attributes - 1) Date of the violation, 2) Location of the violation, 3) Type of violation, 4) Fine amount, 5) Paid or unpaid e-challan (Boolean). Using this data, we compute the total amount of fines paid and owed to the government by computing the sum of all the fines associated with each e-challan in our dataset for paid and unpaid e-challans, respectively. Table \ref{tab:data_stats} provides a quantitative description of our dataset. 

\begin{table}[ht]
\centering
\begin{tabular}{ll} 
  \hline
  \textbf{Description} & \\
  \hline
  Total number of e-challans & $3$,$571$,$341$ \\
  Number of paid e-challans & $1$,$082$,$132$ \\
  Number of unpaid e-challans & $2$,$489$,$209$ \\
  Number of unique vehicles & $1$,$177$,$695$ \\
  \hline
  Date of first e-challan in our dataset & $29$th September 2015 \\
  Date of last e-challan in our dataset & $22$st August 2019 \\  
  \hline
  Total amount owed to the government & $808$,$004$,$725$ \\  
  Total amount paid already & $215$,$574$,$325$ \\
  Total amount unpaid till 22nd August & $592$,$430$,$400$ \\
  \hline
  Number of types of violations & $18$ \\
  Number of unique locations & $135$ \\  
  Number of fine denominations & $28$ \\
 \hline
\end{tabular}
\caption{A brief description of the dataset statistics.}
\label{tab:data_stats}
\end{table}

\subsection{Data Preprocessing}

Due to the presence of multiple cameras in a single location,  two cameras may have the same geographical location but a different name in the location field of the e-challan. Thus, we cleaned the data by updating the location field of all e-challans to a specific geographic location to perform the spatial analysis. The data after the cleaning process had $135$ unique locations across the city and was used for further analysis. 

\section{Characterizing users and types of violations}
\label{sec:characterizingusers}

In this section, we characterize the user behaviour with respect to e-challan payment and repeat offences. We also analyze the distribution of different types of violations in our dataset and discuss their spatial plots later. 

\begin{figure}[h!]
    \centering
    \includegraphics[scale=0.30]{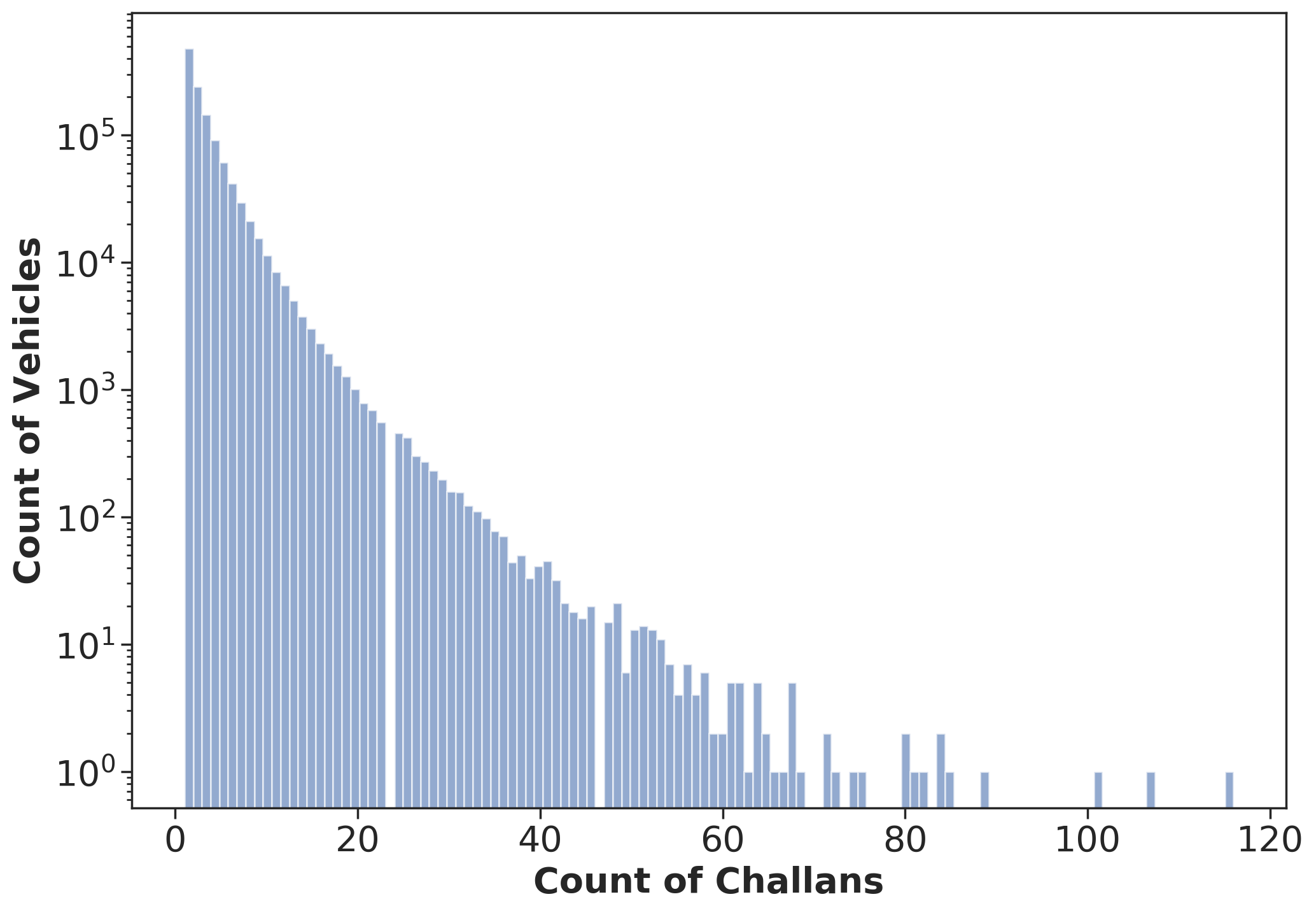}
    \caption{Count of people who have committed specific number of violations. It shows that large number of vehicles have same number of challans.}
    \label{fig:countanalysis}
\end{figure}

\subsection{Repeat offences}

Figure \ref{fig:countanalysis} shows the distribution of users and the number of e-challans issued to each one of them. We compute the median number of e-challans issued to all the users, and this comes out to be around $2$. Thus, at least half the individuals in our dataset have more than $2$ e-challans issued to them. This, suggests that the users are prone to commit traffic violations repeatedly. Figure \ref{fig:countanalysis} also reveals that there are users with even $80$ or more e-challans, suggesting that many of them might not be aware of the e-challans being issued to them. This is further confirmed by the fact that out of the $11$ users with more than $80$ e-challans in our dataset, $5$ of them have not paid even one of their e-challans. 

\begin{table}[ht]
\centering
\begin{tabular}{ll} 
  \hline
  \textbf{Violation Type} & Number of e-challans \\
  \hline
  Red Light Violation & $2$,$102$,$105$ \\
  No Helmet Violation & $1$,$011$,$263$ \\
  Improper Parking & $206$,$765$ \\
  Stop Line Violation & $128$,$658$ \\
  Driving Without Seatbelt & $39$,$202$ \\  
  \hline
  Total Number of e-challans & $3$,$571$,$341$ \\    
  \hline
\end{tabular}
\caption{Distribution of e-challans with violation types.}
\label{tab:violation_stats}
\end{table}

\subsection{Characterizing paid e-challans}
Table \ref{tab:data_stats} shows that the total number of paid e-challans is much less than the unpaid e-challans.  From Table \ref{tab:data_stats}, we see that the ratio of unpaid to paid e-challans is $2.3$, and the ratio of fine amount of the unpaid e-challans to that of paid e-challans is $2.75$. This suggests that the majority of paid e-challans consist of lower fine amounts as the fine payment ratio is $1.19$ times the issued challans ratio. We further analyze the distribution of paid e-challans ratio with respect to the fine amount and find that fines with lesser amounts like Rs. $100$, and $50$ are more likely to be paid as compared to those of higher denominations like $2$,$000$. The difference in the ratio is drastic, with almost $37.37\%$ of the Rs. 100 fines being paid as compared to $15.44\%$ of Rs. 2000 fines. 
\begin{figure}[h!]
    \centering
    \includegraphics[scale=0.44]{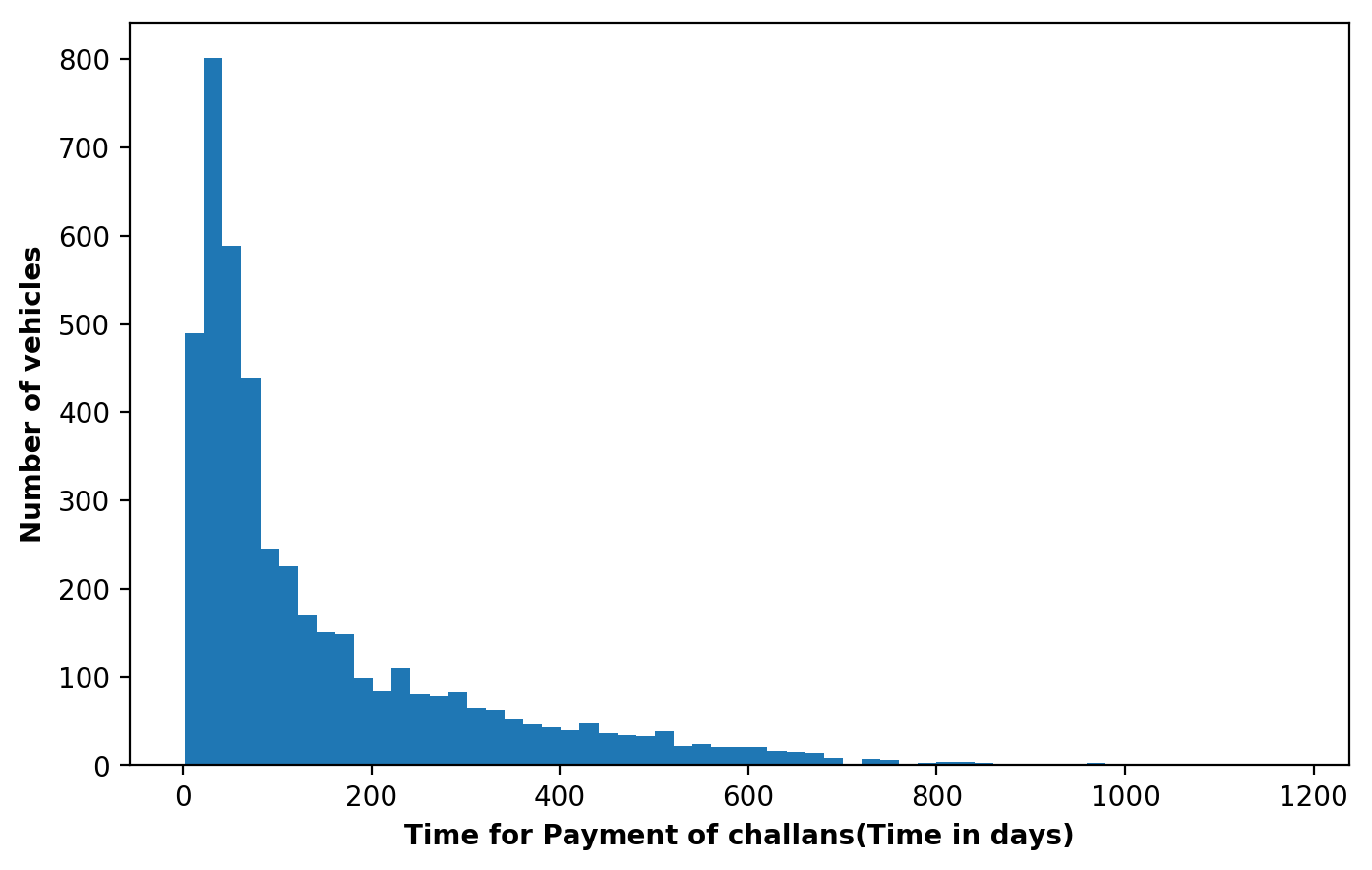}
    \vspace{-1.0em}
    \caption{Count of vehicles with their e-challan payment time.}
    \label{fig:paymenttime}
\end{figure}
In an ideal setting, the distribution of ratio of paid e-challans is expected to be uniform with respect to the type of violation, given that the fine amount is same. However, we find that the e-challans of \textit{overspeeding violation} have drastically higher ratio ($0.60$) of paid e-challans as compared to $0.12$ for \textit{brts lane violation} where their average fine amount is Rs. $1017.67$ and Rs. $1265.2$ respectively.

To characterize user behavior, we analyze the users who had both paid and unpaid e-challans issued to their vehicles. Out of the total vehicles as shown in Section \ref{sec:minidataset}, $229$,$042$ vehicles have both paid and unpaid e-challans issued. Ideally, it is expected that the unpaid e-challans for any vehicle are the recently issued e-challans, but interestingly $31.5\%$ of those vehicles had at least one unpaid e-challan before the last paid e-challan. We characterize these users and find that the fine amounts such as Rs. $50$ contribute much less than the higher denomination fines. In terms of violation types, we find that \textit{no helmet violations} constituted more than $50\%$ of the total e-challans in this category which differs drastically from the normal distribution as shown in Table \ref{tab:violation_stats}. We conclude from this that the violation type and fine amount of the e-challans play a significant role in characterizing fine payment.

To analyze the fine payment behaviour of the users, we compute the average payment time for $4$,$500$ random users sampled from our dataset, as discussed in Section \ref{sec:dataset}. We find that on an average, the users paid back their fines in $153$ days. A histogram of the average fine payment time of each of the $4$,$500$ users is given in Figure \ref{fig:paymenttime}. It indicates that there is a lack of incentive to pay the fines early. A reason for the high average fine payment time could be the lack of awareness amongst the users about the e-challan system, and this needs to be adequately tackled for the system to be effective. 

\begin{figure}[h!]
    \centering
    \includegraphics[scale=0.30]{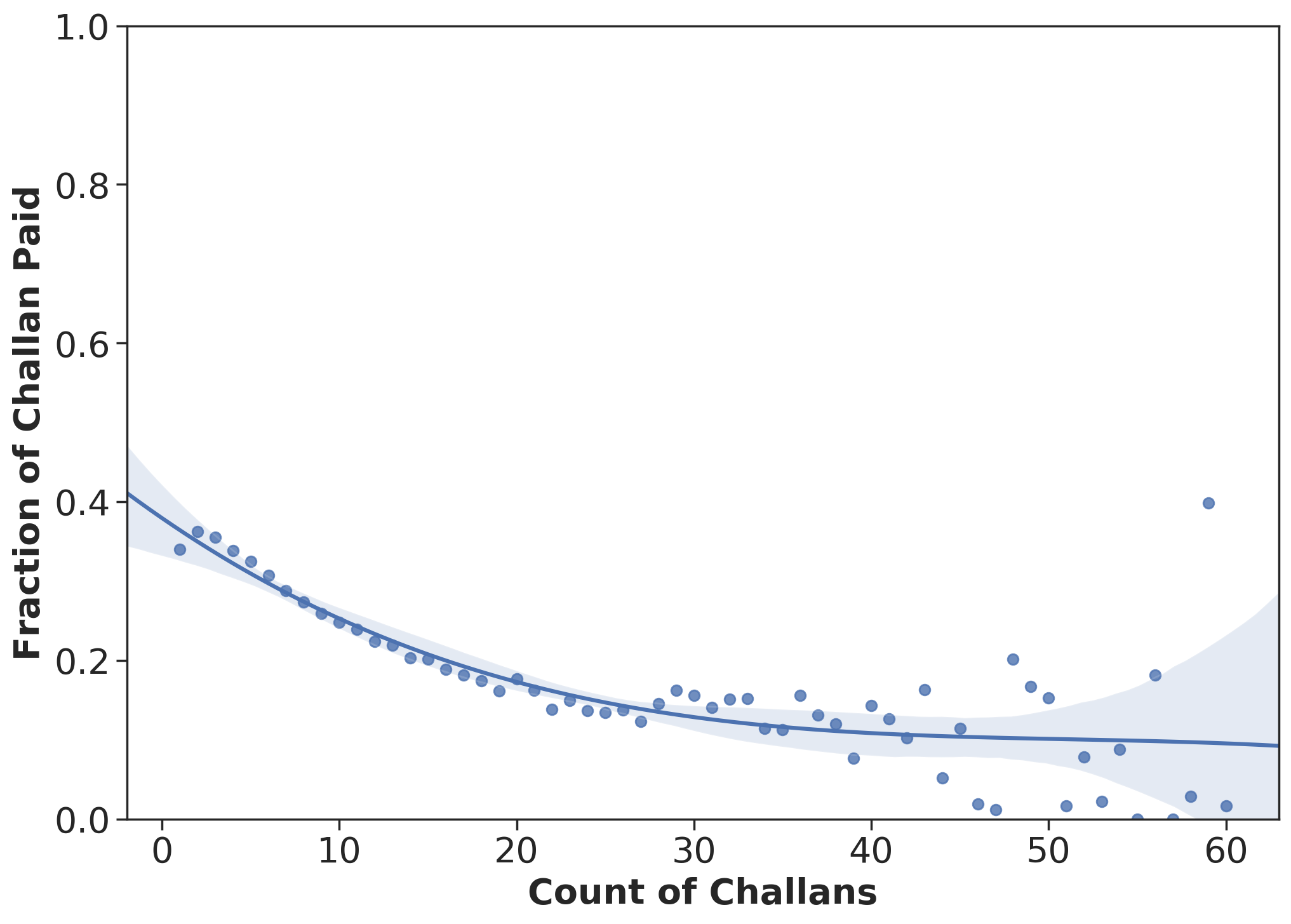}
    \caption{Fraction of e-challans paid by people decreasing with number of violations.}
    \label{fig:payratio}
\end{figure}

We also analyze the impact of the number of e-challans that a user has and their fine payment ratio. We plot the distribution of the number of e-challans issued and the average ratio of e-challan payment in Figure \ref{fig:payratio}. For Figure \ref{fig:payratio}, we removed the data of $42$ users who had an exceptionally high number of violations as they were outliers (less than $0.01\%$ of the users). We can see that people with a lot of e-challans are less likely to pay the fines back as compared to those with fewer e-challans. The distribution fits a third-order curve that decreases with increasing e-challans. The above analysis suggests that the fine payment is highly dependent on the user, the fine amount, and the type of violation associated with the e-challan.

\subsection{Violation types analysis}

The number of e-challans issued for different types of violations is presented in Table \ref{tab:violation_stats}. We present the numbers for the top $5$ violation types in our dataset that account for more than $1\%$ of all the e-challans. We can infer from Table \ref{tab:violation_stats} that the top $2$ violations by themselves account for more than $87\%$ of all the e-challans issued in our dataset. Thus, specific targeted measures can be taken to reduce the number of such violations committed as compared to general measures that target all violations equally. 

\begin{figure}[h]
    \centering
    \includegraphics[width=0.49\linewidth,height=3.2cm]{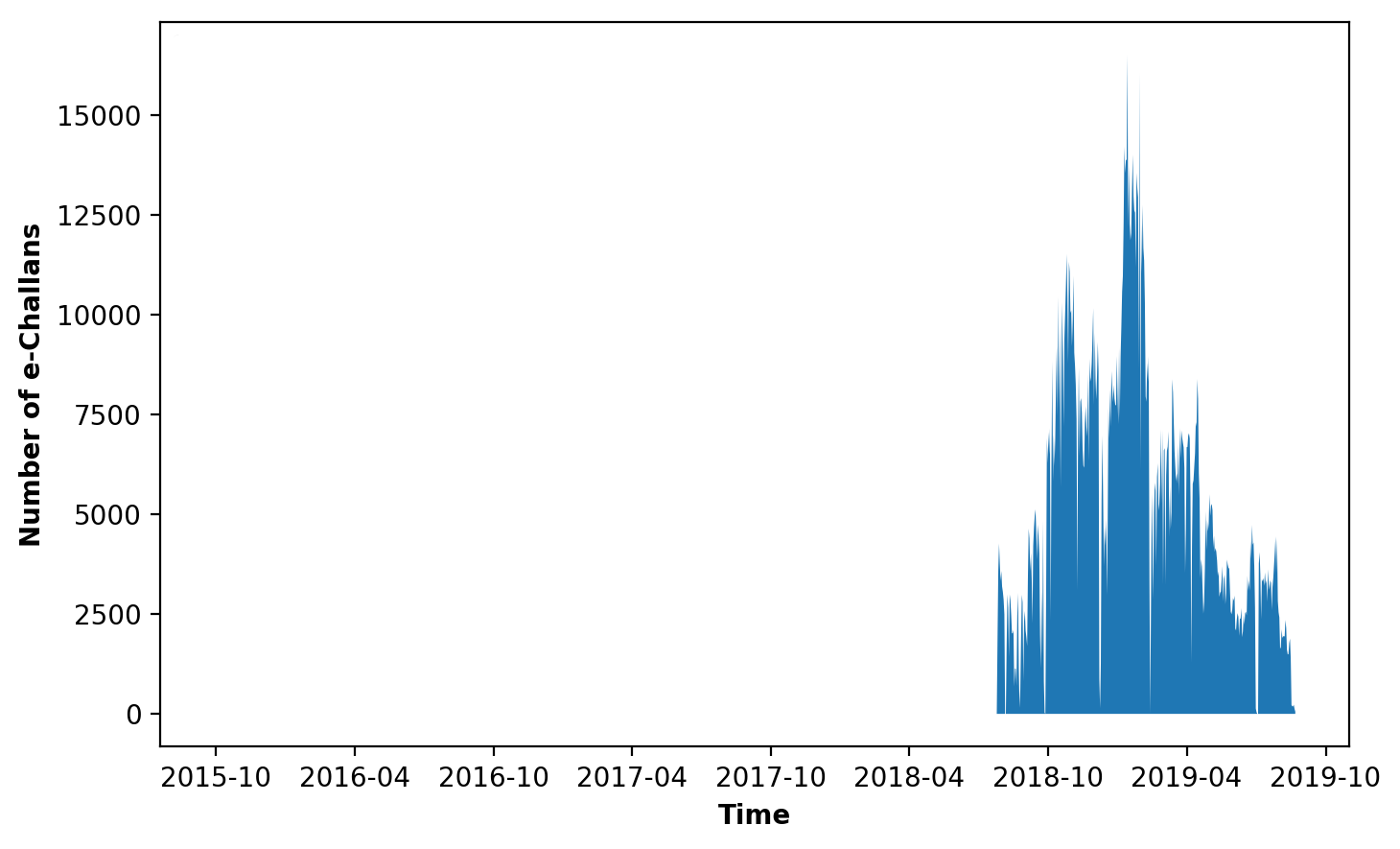}
    \includegraphics[width=0.49\linewidth,height=3.2cm]{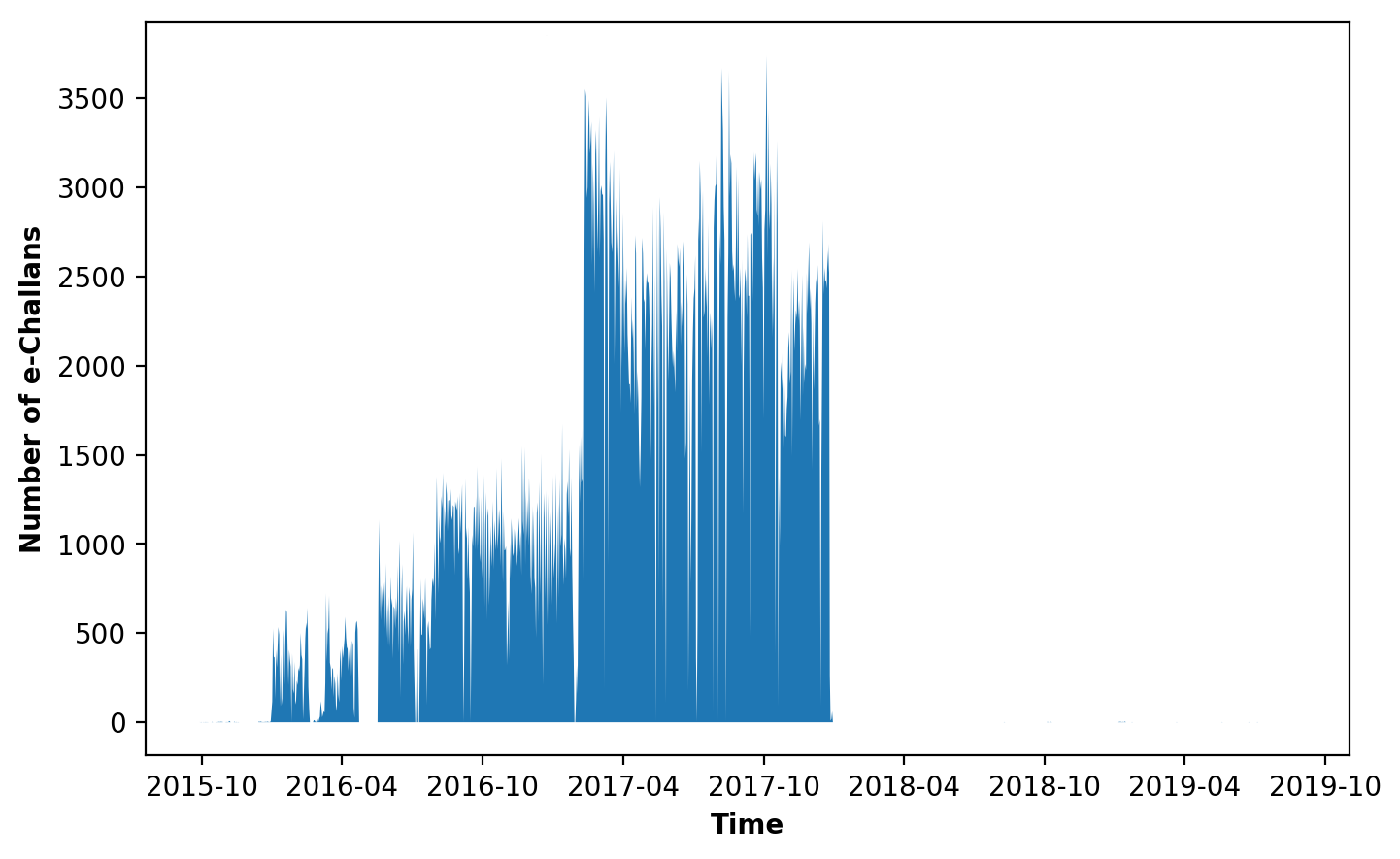}
    \caption{Timeseries plots for \textit{red light violation} (left) and \textit{wearing no helmet on two-wheeler violation} (right) in our dataset.}
    \label{fig:violation_timeseries}
\end{figure}

We analyze the violation types from a temporal perspective and show in Figure \ref{fig:violation_timeseries} that \textit{no helmet violations} are more dominant in the first half of the time-series, which drastically drops after $2017$. On the other hand, \textit{red light violations} are prevalent from the later half of $2018$. Similar temporal bias is observed for other violation types as well, which raises questions on the efficiency of the system and motivation behind it.

\begin{figure}[h]
    \centering
    \setlength{\fboxrule}{1pt}
    \framebox{\includegraphics[width=0.46\linewidth,height=3cm]{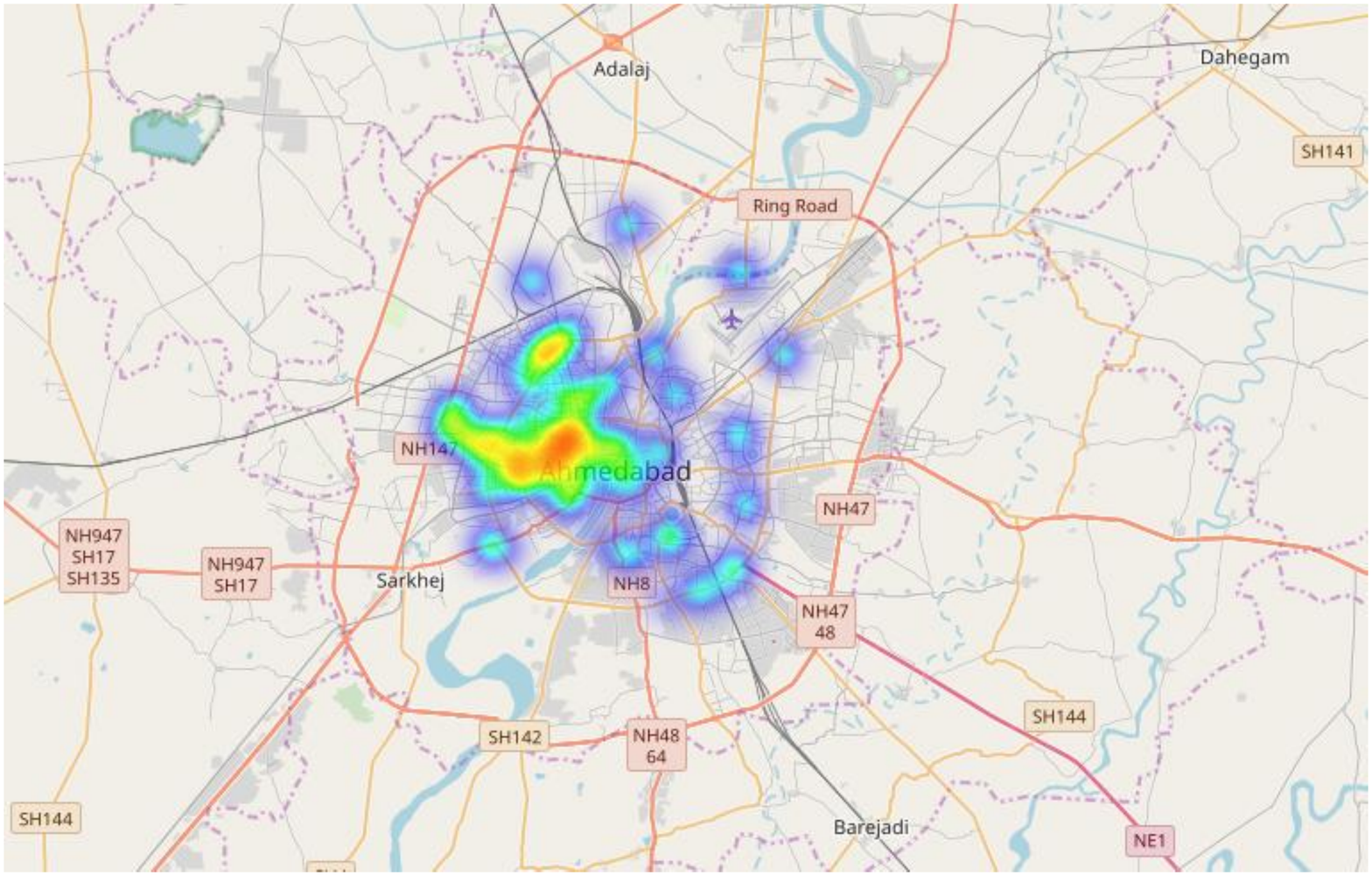}}
    \framebox{\includegraphics[width=0.46\linewidth,height=3cm]{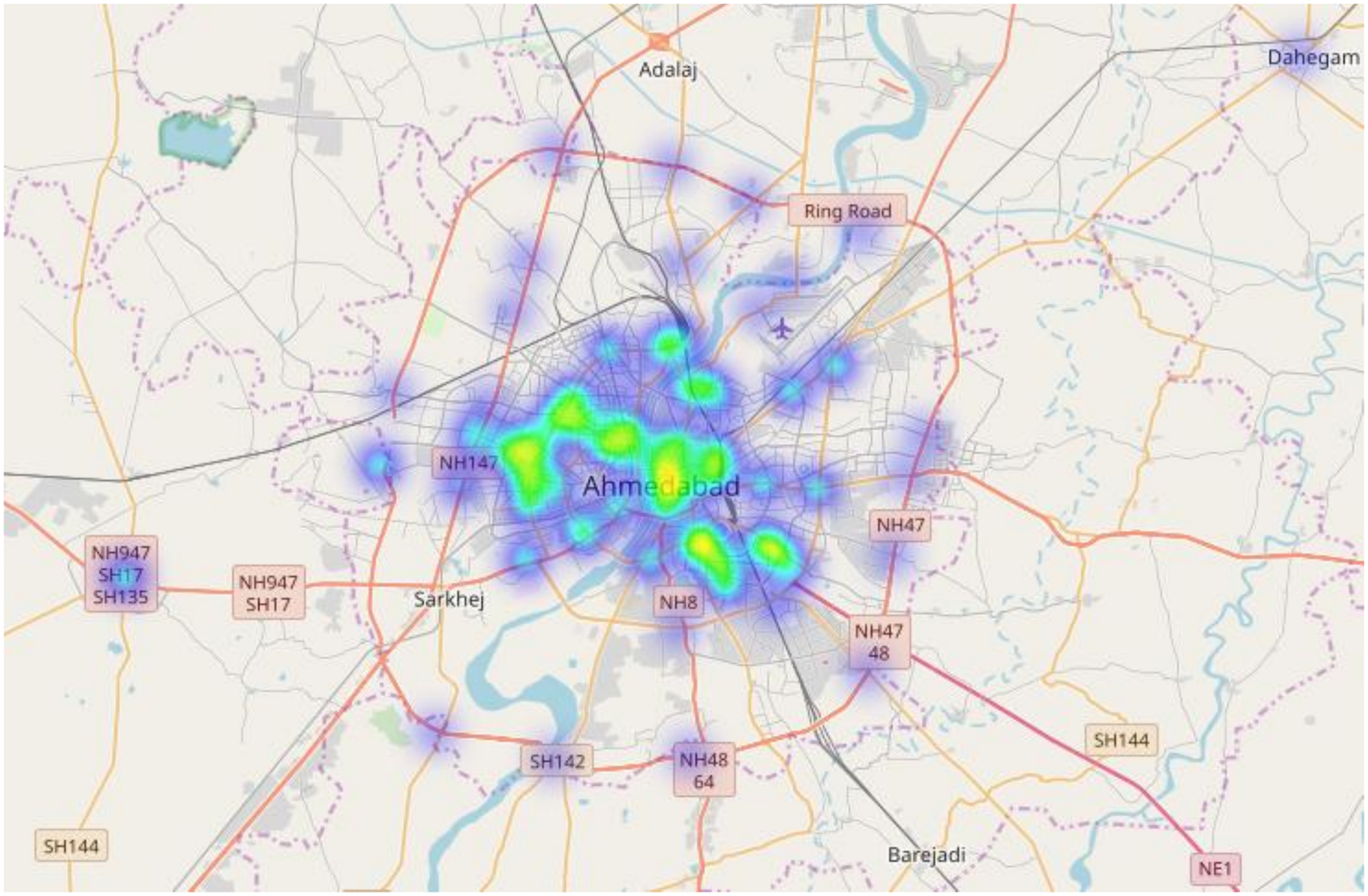}}
    \caption{Heatmaps for \textit{red light violation} (left) and \textit{wearing no helmet on two-wheeler violation} (right) in our dataset.}
    \label{fig:violation_heatmaps}
\end{figure}

We also analyze the violation types from a spatial perspective. We can see in Figure \ref{fig:violation_heatmaps} that the hot-spots of different violation type varies according to the location. From Figure \ref{fig:violation_heatmaps}, we see that \textit{no helmet violations} are more widely distributed across the city as compared to \textit{red light violations} which are concentrated in few regions. Most of the red light violations in our dataset occurred in two specific areas - Navarangpura, a university region and Dariyapur, which lies near the railway station. The implication of the above analysis being, location of a violation type is a specific function of that violation type and are likely to be clustered at certain regions of the city.

\begin{figure*}[!t]
\centering
    \setlength{\fboxrule}{1pt}
    \subfigure[]{\framebox{\includegraphics[width=3.7cm, height=2.7cm]{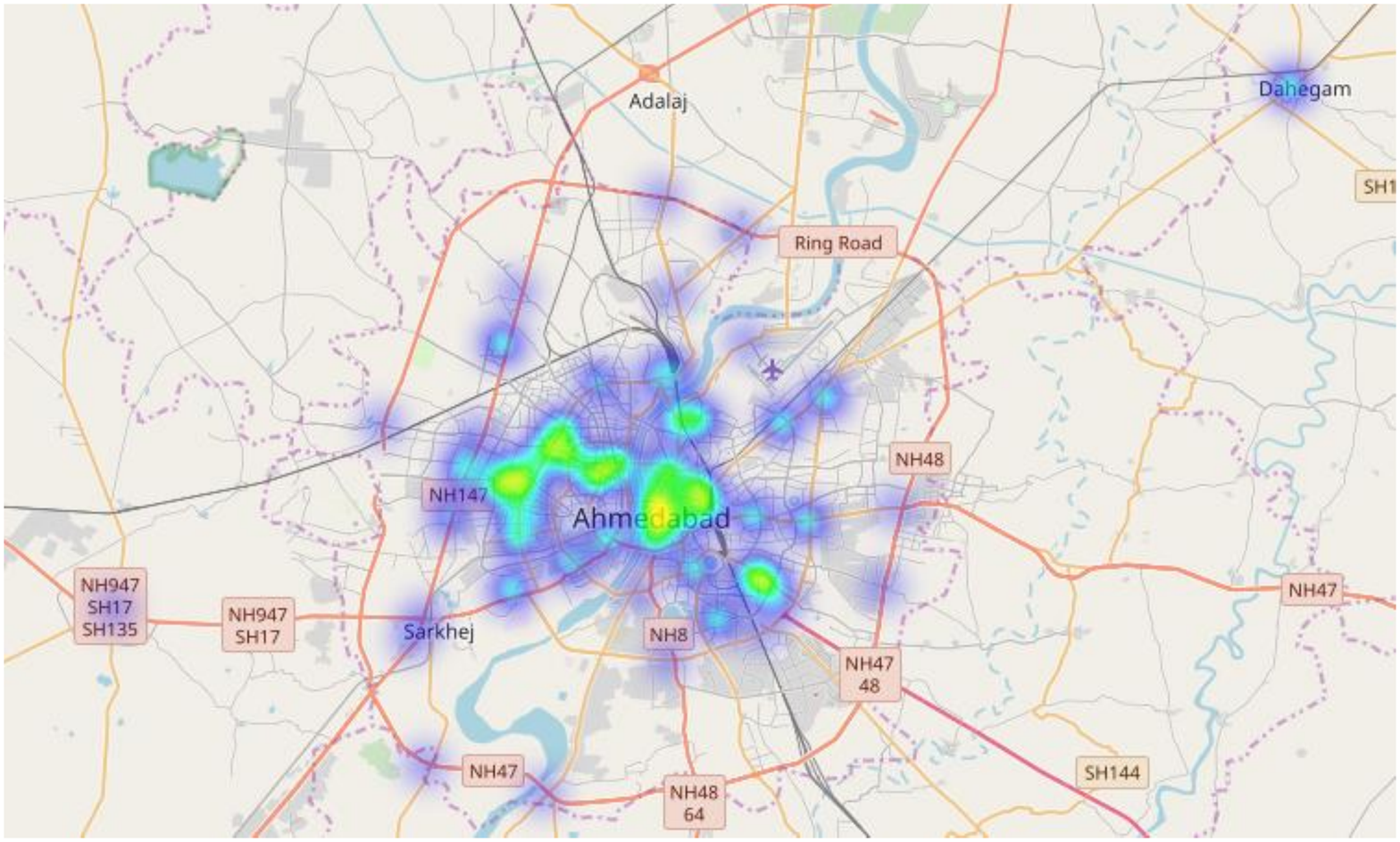}}}
    \hspace{0.5em}
    \subfigure[]{\framebox{\includegraphics[width=3.7cm, height=2.7cm]{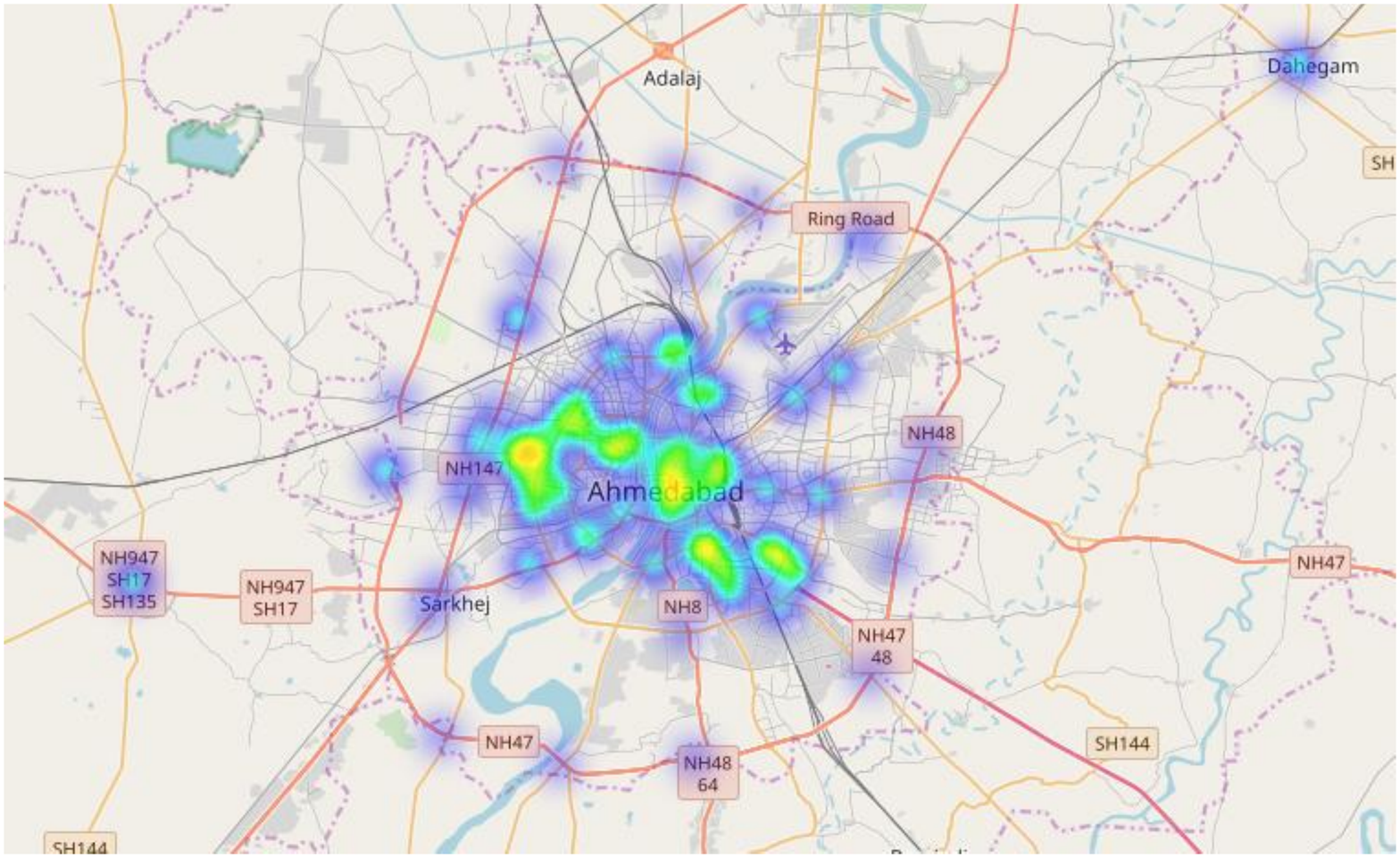}}}
    \hspace{0.5em}    
    \subfigure[]{\framebox{\includegraphics[width=3.7cm, height=2.7cm]{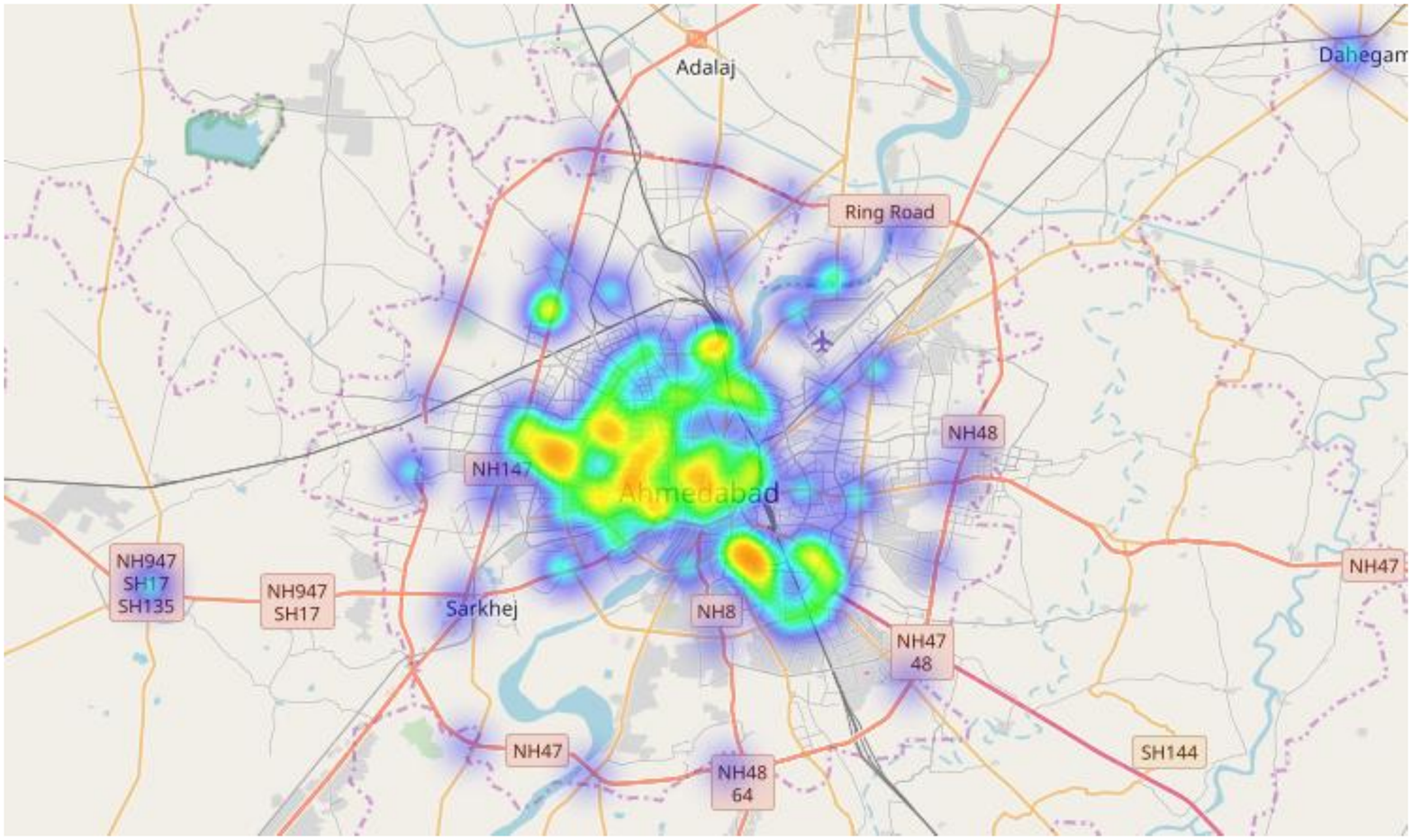}}}
    \hspace{0.5em}    
    \subfigure[]{\framebox{\includegraphics[width=3.7cm, height=2.7cm]{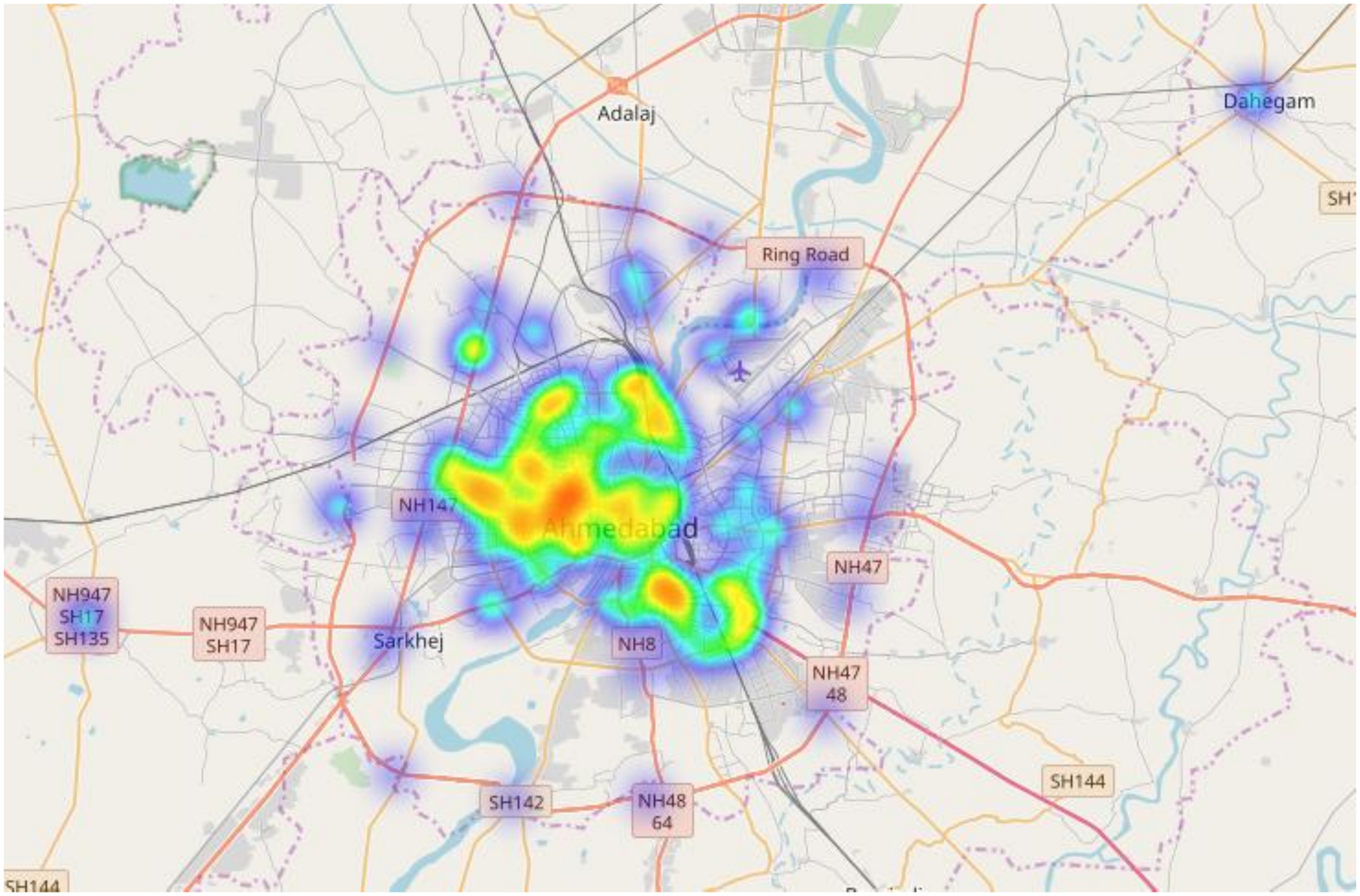}}}
\caption{Heat map of e-challan progression from 2016 to 2019. The figures shows e-challan distribution till year 2016, 2017, 2018, and 2019 in the sub-figures (a), (b), (c), and (d) respectively. It shows clearly how different regions emerge over time along with the increase of e-challans in some regions.}
\label{fig:heatmap}
\end{figure*}

\begin{figure*}[!t]
    \includegraphics[width=0.7\linewidth, height=7.5cm]{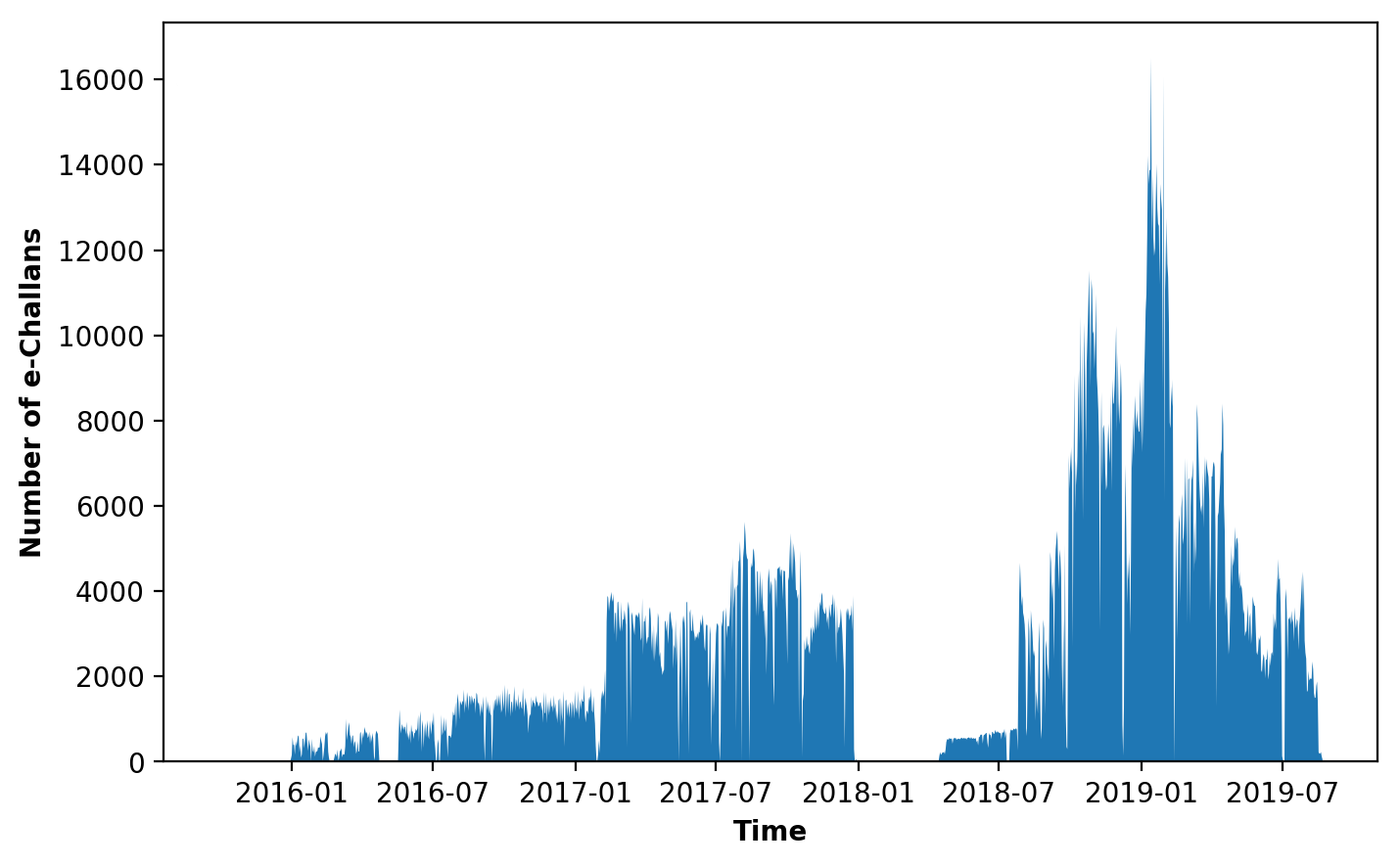}
    \caption{The time series plot of all the e-challans in our dataset. There is a steep increase/decrease in the number of e-challans issued during festival days.}
    \label{fig:temporal-plot}
\end{figure*}

\section{Characterizing Temporal and Spatial Patterns}
\label{sec:spatiotemporal}
In this section, we characterize the spatial and temporal patterns that emerge from our data. Spatial analysis is useful to understand the presence of large as well as a small cluster of locations that account for most of the traffic violations. Temporal analysis, on the other hand, allows us to analyze the general trends in the occurrence of traffic violations and when it is more prevalent. 

\begin{table}[ht]
\centering
\begin{tabular}{ll} 
  \hline
  \textbf{Location} & Number of e-challans \\
  \hline
  Shyamal & $232$,$149$ \\
  Paldi & $218$,$555$ \\
  Vijay Cross Roads & $150$,$358$ \\
  Shashtrinagar & $149$,$631$ \\
  Sardar Patel Statue & $147$,$743$ \\  
  \hline
  Total Number of e-challans & $3$,$571$,$341$ \\  
  \hline
\end{tabular}
\caption{Distribution of e-challans with location.}
\label{tab:location_stats}
\end{table}

\subsection{Spatial analysis}
We use spatial analysis to investigate the presence of certain hotspots where traffic violations are more likely. In Table \ref{tab:location_stats}, we show that the top 5 locations in terms of the number of e-challans issued account for approximately $25\%$ of all the e-challan data collected. \textit{Shyamal} region had the most number of e-challans in our dataset and by itself accounted for approximately $7\%$ of all the e-challans in our dataset. Thus, the data reveals that most of the traffic violations are concentrated in only a few regions of the city.

To locate the smaller regions which also account for a lot of e-challans, we plot the heatmaps of the traffic violations across the city from 2016 to 2019 collectively as shown in Figure \ref{fig:heatmap}. We can infer from Figure \ref{fig:heatmap} (d) that traffic violations are concentrated in a few regions across the city. Also, most of the violations occurred in the central regions of the city and were more concentrated on the left side of the \textit{Sabarmati river}. These insights can be used to take targeted intervention measures for different regions of the city. From Figure \ref{fig:heatmap}, we note that the e-challan system in Ahmedabad has been progressively covering more regions every year, which shows the efficiency of this system. 

\subsection{Temporal Analysis}
Temporal analysis of the data allows us to find patterns about the occurrence of traffic violations in general. Figure \ref{fig:temporal-plot} shows the distribution of all the e-challans issued for $4$ years. It has certain empty regions between January and April 2018 because we do not have data of that period \cite{echallandown}. One trend that can be discerned from Figure \ref{fig:temporal-plot} is that the number of e-challans is continuously increasing over the years with the maximum number of e-challans issued during January 2019. We also analyze the plots to see if there are any spikes around the festival days. In general, there is a steep drop or increase in the number of e-challans issued during festival days. We observe that $2-3$ days before Rath Yatra (Chariot Procession) - July 14, 2018, and July 4, 2019 - the number of e-challans issued is zero as the police personnel were on security duty. This suggestion is further strengthened by the fact that during a few other festivals like Muharram, Eid-ul-Fitr and Diwali, there is a dip in the number of e-challans issued. However, on some other popular festivals in Gujarat, such as Navratri, Rakshabandhan, Janmashtmi and Ganesh Chaturthi, there is a notable rise in the number of e-challans issued. The highest number of e-challans ($16$,$500$) issued on a single day in our dataset was on January 13, 2019, which is a day before Makar Sakranti, one of the most widely celebrated festivals in Ahmedabad. The underlying trend of the high number of e-challans issued a few days before certain festivals continues during the day of the festival as well. 

\section{Predicting Recidivism}
\label{sec:recidivism}
The analysis in previous sections describes in detail the presence of users who repeat offences several times. There have been several papers that have tried to predict the instances of recidivism amongst criminals, youth convicts and sex offenders\cite{ritter2013predicting, hanson1998predicting}. Similar to some previous approaches, we also identify factors strongly related to recidivism. We use machine learning to train a binary classifier that predicts recidivism amongst users. We experiment with multiple types of classification algorithms and perform detailed ablation on multiple combinations of the features. We first describe the dataset that we used to estimate recidivism, followed by the experiments on several classification models and ablation studies. 

\subsection{Recidivism dataset}
To get the ground truth data for recidivism, we consider all the people who had an e-challan before April 2019 and assign them a label for recidivism. If a person received another e-challan in our dataset after April 2019, then we consider it to be a case of recidivism and hence assign it the corresponding label. 
In our dataset,  $636$,$482$ people had been issued an e-challan before April 2019. Of these people, $468$,$731$ of them had committed a violation after this date.  

We thus model our problem as a two-class classification problem with $468$,$731$ positive samples (people who repeat violations) and $167$,$751$ negative samples (people who do not repeat violations).

\subsection{Features}
We identify several features based on a person's past violation history that could be useful for predicting recidivism. We provide a detailed description of these features and their importance below.

\begin{enumerate}
    \item \textbf{Number of paid e-challans (Paid)}: We compute the number of paid e-challans that each person had before April 1st and use it as a feature for our model. As discussed in Section \ref{sec:characterizingusers}, there exists the possibility of people being unaware of the e-challans being issued to them. Thus, payment of the fine amount for an e-challan implies that the user is aware of the fines being levied and would be less likely to commit more traffic violations. 
    \item \textbf{Number of unpaid e-challans (Unpaid)}: The number of unpaid e-challans along with the number of paid e-challans gives us the total number of e-challans issued to a person. A person with a very high number of e-challans issued is more likely to re-offend as compared to someone who has less number of violations. 
    \item \textbf{Mean time difference between consecutive violations (Frequency)}: This metric provides the frequency/regularity in which a person commits traffic violations. Let $D_i$ and $D_{i+1}$ represent the days on which a person received e-challan number $i$ and $i+1$ respectively. Also, $T$ is the total number of violations for a given person. We compute the mean time difference between consecutive violations ($A$) as:
    \begin{equation*}
        A = \frac{\sum_{i=1}^{i=T-1} |D_{i+1} - D_i|}{T - 1} 
    \end{equation*}
    Lower values for this metric signify that a person violates traffic rules very frequently, thus increasing their likelihood of violating again. Similarly, higher values for this metric mean that the person is involved in a traffic violation in long intervals of time only, suggesting that he/she committed the violation by mistake. 
    \item \textbf{Number of days since the last e-challan (Recency)}: A user may have committed a lot of traffic violations a few years ago but may have become more careful since then. Thus, our model should also be conditioned on the time in which a person last committed a traffic violation.
    \item \textbf{Entropy of traffic violation types (Entropy)}: We also compute the entropy of all traffic violation types for each person in our dataset. For each user, we first compute the set $T$ of traffic violation types that a user committed earlier. Let $T_i$ be the number of violations of type $i  \forall i \in T$. Then, for each person, we compute the entropy (E) as: 
    \begin{equation*}
        E = \sum_{i \in T} \left[ \frac{T(i)}{\sum \limits_{i \in T} T_i} \right] \left[ \log{\frac{T(i)}{\sum \limits_{i \in T} T_i}} \right]
    \end{equation*}
    The entropy for each user thus measures the variation in the type of violations that each user committed in the past. A person committing a single type of violation like \textit{Driving without helmet} is more likely to commit the same violation again as compared to someone who has different types of violation. 
\end{enumerate}

We use these features to predict recidivism and also conduct an ablation study on the most important features out of these. One of the main benefits of these features discussed above is that they can be easily computed for a new source of data. These features do not leverage any personally identifiable information like age, gender and religion, and are less likely to be biased by such factors.

\subsection{Classification Techniques}
We experimented with the following classifiers: Logistic Regression, Linear Support Vector Machines (SVM), Multi-Layered Perceptron (MLP), Random Forest\cite{breiman2001random} and XG-Boost \cite{chen2016xgboost}. We present the models and their corresponding features and hyper-parameters below:

\begin{itemize}
    \item \textbf{Logistic Regression: } We train a logistic regression model with L2 regularization. We set the value of $C$ (Inverse of regularization strength) to $1$.
    \item \textbf{Linear SVM: } We train a Support Vector Machine with linear kernel and L2 regularization. We set the value of $C$ to $1$ and train it for a total of $5,000$ iterations.
    \item \textbf{Multi-Layer Perceptron: } We train a multilayer perceptron (Neural Network) with two hidden layers consisting of $32$ and $64$ neurons respectively. We train the network for $X$ epochs and use the Adam \cite{kingma2014adam} optimizer.
    \item \textbf{Random Forest: } We train a random forest classifier on the training data with $200$ base learners and a max depth of the tree fixed at $30$.
    \item \textbf{XG-Boost: } We use XG-Boost to learn the gradient boosted decision tree models. We train it with $147$ base learners,  max depth of $5$ for the trees and a learning rate of $0.28$. 
\end{itemize}

\subsection{Analysis}

To ensure that the results are not a false indication of the performance of the models, we make sure at the time of train-test split that data is properly shuffled. We use $80\%$ of the data for training and the remaining $20\%$ of data for testing purposes. We also report the precision, recall and F1 score apart from testing accuracy to make sure that the insights obtained are correct and report them in Table \ref{table:classifier}.

\begin{table}[ht]
\centering
\begin{tabular}{lcccc}
\toprule
 Classifier & Test & Precision & Recall & F-1 Score  \\ 
 & Accuracy & \\
 \midrule
 Logistic & $0.88$ & $0.88$ & $0.88$ & $0.87$ \\
 Regression & & \\
 Linear SVM & $0.85$ & $0.86$ & $0.85$ & $0.84$ \\
 MLP & $0.94$ & $0.94$ & $0.94$ & $0.94$ \\
 Random Forest & $\mathbf{0.95}$ & $\mathbf{0.95}$ & $\mathbf{0.95}$ & $\mathbf{0.95}$ \\
 XG-Boost & $0.95$ & $0.95$ & $0.95$ & $0.95$ \\
 \bottomrule
\end{tabular}
\caption{Performance of various classification methods on our test dataset.}
\label{table:classifier}
\end{table}

\begin{figure}[!h]
    \centering
    \includegraphics[width=0.8\linewidth]{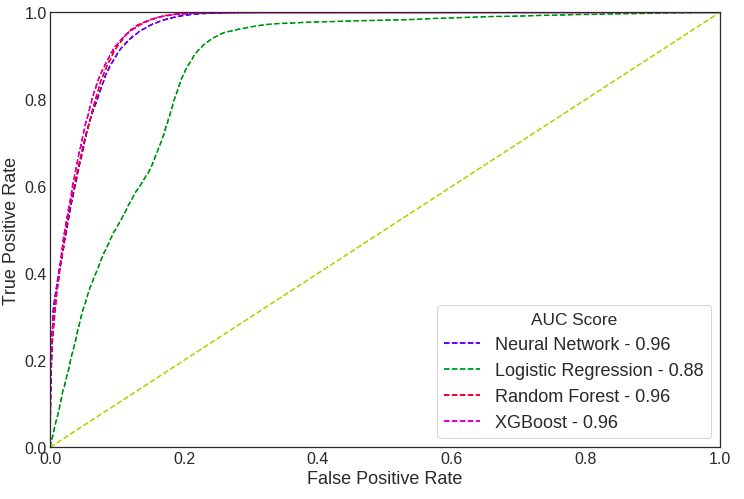}
    \vspace{-1em}
    \caption{AUC of different classification techniques used to predict recidivism of traffic violations. Higher AUC score denotes better classification ability of the model.}
    \label{fig:roc_curve}
\end{figure}

The random forest and XG-Boost classifier perform equally well on our test dataset and have a very high F1 score of $95\%$ and $95\%$ respectively. On the other hand, logistic regression and linear SVM classifiers perform comparatively worse. Figure \ref{fig:roc_curve} shows the receiver operating characteristic (ROC) curve for all the models except the linear SVM, which does not provide any probabilistic estimate. The AUC (Area Under Curve) of random forests and XG-Boost is quite high, suggesting that the classifier can separate the two classes well. 

\subsection{Ablation Study}
We perform ablation studies to find out the features that contribute most to the test accuracy and F1 score. We use random forest classifier to estimate the feature importance as it was one of the best performing models.  The tree-based strategies used by random forest naturally rank by how well they improve the purity of the node. We simply remove one feature at a time from our dataset and train the model on the new set of features and compute its test accuracy. From the results of the ablation study in Table \ref{table:ablation}, we can see that the \textit{entropy} feature has the most impact on the model, followed by \textit{number of unpaid e-challans}. Removing the entropy feature from our training data decreases the test accuracy drastically from $95\%$ to around $76\%$. The Table \ref{table:ablation} also shows that temporal features like \textit{time since the last e-challan} and \textit{mean frequency of e-challans} do not have a significant impact on the test accuracy.

\section{Discussion}
\label{sec:discussion}
We created a dataset of e-challans in the city of Ahmedabad and analyzed the data to gain some insights about traffic violations in the city. We perform spatial analysis to show that there are regions in the city with high intensity of traffic violations. The temporal analysis depicts that the number of e-challans issued show a sharp spike or drop around festival days. We show that few violations types account for most of the e-challans and a significant percentage of people are repeat offenders.  This suggests that we need to take special targeted intervention measures to gauge these cases. We also show that the payment of e-challans is influenced by the fine amount and the type of violation associated with it. We describe the features used to predict recidivism in traffic violations and train a random forest model to predict it. We also carry out detailed ablation study and show that the features related to the type of violations matter more than the temporal features and the absolute number of e-challans. The cornerstone of our work is the detailed analysis of traffic violation data to gain unique insights. These can be used by traffic police and government to make the roads safer.

\begin{table}[ht]
\centering
\begin{tabular}{lcccc}
\toprule
 Feature & Test Accuracy & Precision & Recall & F-1 Score  \\ 
 removed & \\
 \midrule
 \textit{Paid} & $0.88$ & $0.88$ & $0.88$ & $0.88$ \\
 \textit{Unpaid} & $0.86$ & $0.86$ & $0.86$ & $0.86$ \\
 \textit{Frequency} & $0.91$ & $0.92$ & $0.91$ & $0.91$ \\
 \textit{Entropy} & $\mathbf{0.76}$ & $\mathbf{0.74}$ & $\mathbf{0.76}$ & $\mathbf{0.74}$ \\ 
 \textit{Recency} & $0.94$ & $0.94$ & $0.94$ & $0.94$ \\
 \bottomrule
\end{tabular}
\caption{Results of the ablation study on the features. Removal of \textit{entropy} feature affects the model worst.}
\label{table:ablation}
\end{table}

\section{Future Work}
\label{sec:futurework}
The study is currently limited to only one city, and thus the inferences made can not be generalized to other cities across the world. We can perform a similar analysis on traffic violation data of other cities to make bolder and more general claims. The current analysis does not leverage any form of demographic data like gender or age due to lack of such data. Studies on road accident data have shown that the demographics of the people also significantly affect the fatality and the same is true for our analysis. Thus, analyzing traffic violations from the prism of demographics characteristics is another interesting line of work. 

\bibliographystyle{unsrtnat}
\bibliography{references}

\end{document}